\newcommand\apjcls{1}
\newcommand\aastexcls{2}
\newcommand\othercls{3}

\newcommand\papercls{\aastexcls}
\documentclass[tighten,times,twocolumn]{aastex62}  

\if\papercls \apjcls
\usepackage{apjfonts}
\else\if\papercls \othercls
\usepackage{epsfig}
\usepackage{margin}
\usepackage{times}
\fi\fi
\usepackage{ifthen}
\usepackage{natbib}
\usepackage{bm}
\usepackage{amssymb, amsmath}
\usepackage{appendix}
\usepackage{etoolbox}
\usepackage[T1]{fontenc}
\usepackage{paralist}
\usepackage{newtxtext,newtxmath}
\if\papercls \apjcls
\newcommand\aas{\ref@jnl{AAS Meeting Abstracts}}
\newcommand\dps{\ref@jnl{AAS/DPS Meeting Abstracts}}
\newcommand\maps{\ref@jnl{MAPS}}
\else\if\papercls \othercls
\usepackage{astjnlabbrev-jh}
\fi\fi

\if\papercls \aastexcls
\hypersetup{citecolor=blue, 
            linkcolor=blue, 
            menucolor=blue, 
            urlcolor=blue}  
\else
\usepackage[
bookmarks=true,           
bookmarksnumbered=true,   
colorlinks=true,          
citecolor=blue,           
linkcolor=blue,           
menucolor=blue,           
urlcolor=blue,            
linkbordercolor={0 0 1},  
pdfborder={0 0 1},
frenchlinks=true]{hyperref}
\fi
\if\papercls \othercls

\else

\fi

\providecommand{\adsurl}[1]{\href{#1}{ADS}}

\makeatletter
\patchcmd{\NAT@citex}
  {\@citea\NAT@hyper@{%
     \NAT@nmfmt{\NAT@nm}%
     \hyper@natlinkbreak{\NAT@aysep\NAT@spacechar}{\@citeb\@extra@b@citeb}%
     \NAT@date}}
  {\@citea\NAT@nmfmt{\NAT@nm}%
   \NAT@aysep\NAT@spacechar\NAT@hyper@{\NAT@date}}{}{}

\patchcmd{\NAT@citex}
  {\@citea\NAT@hyper@{%
     \NAT@nmfmt{\NAT@nm}%
     \hyper@natlinkbreak{\NAT@spacechar\NAT@@open\if*#1*\else#1\NAT@spacechar\fi}%
       {\@citeb\@extra@b@citeb}%
     \NAT@date}}
  {\@citea\NAT@nmfmt{\NAT@nm}%
   \NAT@spacechar\NAT@@open\if*#1*\else#1\NAT@spacechar\fi\NAT@hyper@{\NAT@date}}
  {}{}
\makeatother

\makeatletter
\DeclareRobustCommand{\lowcase}[1]{\@lowcase#1\@nil}
\def\@lowcase#1\@nil{\if\relax#1\relax\else\MakeLowercase{#1}\fi}
\makeatother

\DeclareSymbolFont{UPM}{U}{eur}{m}{n}
\DeclareMathSymbol{\umu}{0}{UPM}{"16}
\let\oldumu=\umu
\renewcommand\umu{\ifmmode\oldumu\else\math{\oldumu}\fi}

\if\papercls \othercls

\else

\fi

\let\oldsim=\sim
\renewcommand\sim{\ifmmode\oldsim\else\math{\oldsim}\fi}
\let\oldpm=\pm
\renewcommand\pm{\ifmmode\oldpm\else\math{\oldpm}\fi}
\newcommand\by{\ifmmode\times\else\math{\times}\fi}

\newbox{\wdbox}
\renewcommand\c{\setbox\wdbox=\hbox{,}\hspace{\wd\wdbox}}
\renewcommand\i{\setbox\wdbox=\hbox{i}\hspace{\wd\wdbox}}

\newcount\timect
\newcount\hourct
\newcount\minct
\newcommand\now{\timect=\time \divide\timect by 60
         \hourct=\timect Cltiply\hourct by 60
         \minct=\time \advance\minct by -\hourct
         \number\timect:\ifnum \minct < 10 0\fi\number\minct}

\catcode`@=11

\newcommand\comment[1]{}

\newcommand\commenton{\catcode`\%=14}

\renewcommand\math[1]{$#1$}
\newcommand\mathshifton{\catcode`\$=3}

\let\atab=&
\newcommand\atabon{\catcode`\&=4}

\let\oldmsp=\sp
\let\oldmsb=\sb
\def\sp#1{\ifmmode
           \oldmsp{#1}%
         \else\strut\raise.85ex\hbox{\scriptsize #1}\fi}
\def\sb#1{\ifmmode
           \oldmsb{#1}%
         \else\strut\raise-.54ex\hbox{\scriptsize #1}\fi}
\newbox\@sp
\newbox\@sb
\def\sbp#1#2{\ifmmode%
           \oldmsb{#1}\oldmsp{#2}%
         \else
           \setbox\@sb=\hbox{\sb{#1}}%
           \setbox\@sp=\hbox{\sp{#2}}%
           \rlap{\copy\@sb}\copy\@sp
           \ifdim \wd\@sb >\wd\@sp
             \hskip -\wd\@sp \hskip \wd\@sb
           \fi
        \fi}
\def\msp#1{\ifmmode
           \oldmsp{#1}
         \else \math{\oldmsp{#1}}\fi}
\def\msb#1{\ifmmode
           \oldmsb{#1}
         \else \math{\oldmsb{#1}}\fi}

\def\supon{\catcode`\^=7}

\def\subon{\catcode`\_=8}

\def\supsubon{\supon \subon}

\newcommand\actcharon{\catcode`\~=13}

\newcommand\paramon{\catcode`\#=6}

\comment{And now to turn us totally on and off...}

\newcommand\reservedcharson{ \commenton  \mathshifton  \atabon  \supsubon 
                             \actcharon  \paramon}

\catcode`@=12
\reservedcharson

\if\papercls \apjcls

\else

\fi

\newcommand\chisq{\ifmmode{\chi\sp{2}}\else\math{\chi\sp{2}}\fi}
\newcommand\redchisq{\ifmmode{ \chi\sp{2}\sb{\rm red}}
                    \else\math{\chi\sp{2}\sb{\rm red}}\fi}
\newcommand\Teq{\ifmmode{T\sb{\rm eq}}\else$T$\sb{eq}\fi}
\newcommand\mjup{\ifmmode{M\sb{\rm Jup}}\else$M$\sb{Jup}\fi}
\newcommand\rjup{\ifmmode{R\sb{\rm Jup}}\else$R$\sb{Jup}\fi}
\newcommand\msun{\ifmmode{M\sb{\odot}}\else$M\sb{\odot}$\fi}
\newcommand\rsun{\ifmmode{R\sb{\odot}}\else$R\sb{\odot}$\fi}
\newcommand\mearth{\ifmmode{M\sb{\oplus}}\else$M\sb{\oplus}$\fi}
\newcommand\rearth{\ifmmode{R\sb{\oplus}}\else$R\sb{\oplus}$\fi}

\renewcommand{\bar}[1]{\overline{#1}}

\renewcommand{\bm}[1]{{\mbox{{\boldmath$#1$}}}}	

\shorttitle{Semiconvection in Stars and Planets}

\begin{document}

\title{3D Simulations of Semiconvection in Spheres: Turbulent Mixing and Layer Formation}

\author{J. R. Fuentes}
\affiliation{\rm Department of Applied Mathematics, University of Colorado Boulder, Boulder, CO 80309-0526, USA}

\begin{abstract}
    Semiconvection occurs in regions of stars and planets that are unstable to overturning convection according to the Schwarzschild criterion, yet stable according to the Ledoux criterion. Previous simulations in Cartesian boxes have advanced our understanding of the semiconvective instability, layer formation, and transport properties. However, much less is known about semiconvection in spherical geometry and under the influence of rotation or magnetic fields. We present 3D simulations of semiconvection in the full sphere (including $r=0$), and accounting for rotation. We find that the formation and evolution of semiconvective layers in nonrotating spheres occurs in a similar way to nonrotating Cartesian boxes, in the sense that the critical density ratio at which layers are expected to form is approximately the same in both geometries. Layers rapidly merge once they form, ultimately leading to a fully mixed convective sphere. The transport properties measured through the Nusselt numbers and the buoyancy flux ratio are also similar to results from previous studies in boxes. When rotation is added to the system, layer formation and evolution proceeds in a similar fashion to the nonrotating runs. However, rotation hampers the radial transport of heat and composition, and, as a result, the time required for the sphere to become fully mixed gets longer as the flow becomes more rotationally constrained. We also find that semiconvective layers exhibit spherical mixing in nonrotating cases, whereas in rotating cases, the mixing becomes more cylindrical. We discuss what is needed for future work to build more realistic models.
\end{abstract}


\section{Introduction}
Whether a fluid is stably stratified or unstable to convective overturn plays a crucial role in its dynamics and transport properties. This is particularly true in the interiors of stars and planets, where heat may be transported by convection or by radiation, and which of the two mechanisms is effective determines the evolution of the object. Of particular interest are regions that are unstably stratified in temperature (i.e. unstable to overturning convection according to the Schwarzschild criterion) but stably stratified by a composition gradient (i.e., the fluid is overall stable to overturning convection according to the Ledoux criterion). 
In this situation, the complex interaction between advection and diffusion of heat and chemical species can trigger an instability that takes the form
of overstable gravity waves, driving the fluid to a regime known as double-diffusive convection or semiconvection \citep[see, e.g.,][]{Schwar1958,Kato1966,Spiegel1969}. This process enhances the transport of heat and composition within a fluid with respect to just diffusion, and thus plays a crucial role in shaping the internal structure, cooling history, and magnetism of stars and planets. 

For example, in intermediate-mass and high-mass stars, nuclear burning gradually creates a mean molecular weight gradient at the edge of the convective core. When the gradient becomes strong enough to stabilize the
fluid, the fully convective core shrinks in size. This process leaves behind a semiconvective region, where double-diffusive processes enhance the transport of energy and chemical species between the core and the envelope, which provides additional fuel for nuclear reactions and prolongs the star's main-sequence phase \citep[see, e.g.,][]{Ledoux1947,Tayler1954,Langer1991,Merryfield1995}. Another example is in the context of giant planets. The formation and evolution models of Jupiter and Saturn predict extended gradients of heavy elements immersed in a hydrogen-helium evelope, with the abundance of heavy elements decreasing toward the planet's surface \citep[e.g.,][]{Muller2020,stevenson_et_al_2022}. This creates a stabilizing compositional gradient, which may lead to semiconvection. In this context, semiconvection has been invoked to explain the abnormally large radius of some giant exoplanets \citep{Chabrier_Baraffe_2007}, and the anomalous luminosities of Saturn and Uranus \citep{Leconte_and_Chabrier_2013,Vazan2020}.

Advances in high-performance computing have made it
feasible to study semiconvection in astrophysical fluids using 3D numerical simulations in Cartesian boxes. The seminal calculations of  \cite{Rosenblum2011} and \cite{Mirouh2012} demonstrated that the double-diffusive instability leads to either a state of ``homogeneous oscillatory convection'' where diffusive transport is only modestly enhanced by small scale turbulence, or to a state of ``layered convection'', with numerous, small convective layers separated by thin, diffusive interfaces corresponding to discontinuities in the density of the fluid. The specific state the fluid falls into depends primarily on three non-dimensional parameters: the Prandtl number, the diffusivity ratio, and the density ratio, defined respectively as

\begin{equation}
\mathrm{Pr} = \dfrac{\nu}{\kappa_T},\quad \tau = \dfrac{\kappa_C}{\kappa_T},\quad R_{\rho} = \dfrac{N^2_C}{N^2_T}, \label{eq:parameters}
\end{equation}
where $\nu$ is the kinematic viscosity of the fuid, $\kappa_C$ and $\kappa_T$ are the
compositional and thermal diffusivities, respectively, and $N^2_T$ and $N^2_C$ are the Brunt–Väisälä frequencies due to the thermal and compositional stratification, respectively. 
The instability gets excited when 
\begin{equation}
 1<R_{\rho} < \dfrac{\rm Pr + 1}{\rm Pr + \tau}.
\end{equation}
Within this range, spontaneous formation of multiple convective layers is observed for $R_{\rho} \in [1, R_L]$, while for $R_{\rho} \in [R_L, (\mathrm{Pr} + 1)/(\mathrm{Pr} + \tau)$, the instability leads to a state of homogeneous semiconvection. $R_L$ corresponds to the point where the composition-to-heat buoyancy flux ratio reaches a minimum with respect to $R_{\rho}$ \citep[see, e.g.,][]{Radko2003,Mirouh2012}. In stars and planets, both $\mathrm{Pr}$ and $\tau$ are $\ll 1$, allowing the double-diffusive instability to develop across a wide range of stratifications $R_\rho$. However, the exact value of the critical density ratio $R_L$, which determines the onset of layer formation, depends on the fluid and flow properties and is difficult to estimate from first principles.

Although further numerical studies have investigated the transport properties of double-diffusive convection \citep{Wood2013,Moll2016} and the layer formation process under different boundary conditions \citep{Zaussinger2019,Fuentes2022,Tulekeyev2024}, much less is known about semiconvection under the effects of additional physics such as magnetic fields and rotation. \cite{Sangui2022} investigated double-diffusive convection in the presence of a uniform vertical background magnetic field, finding that a sufficiently strong field delays layer formation and substantially reduces the internal flux across the layers. Similarly, uniform rotation in the anti-parallel direction to vertical gravity has been shown to produce similar effects to those of magnetic fields \citep{Moll2017_rot,Lian2021,Fuentes2024}. 

In order to better model real astrophysical objects, it is essential to adopt spherical geometry. For example, in a rapidly rotating planet or star, the latitude-dependent misalignment between gravity and the rotation vector can result in anisotropic mixing and heat transport \citep[e.g.,][]{Gastine2023}, thereby influencing the formation and dynamics of semiconvective layers. To our knowledge, the only attempt to study semiconvection in spherical geometry was done by \cite{Blies2014}, who conducted simulations in rotating spherical shells (i.e., with the coordinate singularity at $r = 0$ avoided by making a ``cutout'' in the center of the sphere). Similar to previous studies in Cartesian geometries, they found that strong rotation suppresses the transport of both temperature and composition. However, it remains unclear from their results and displayed flow morphologies whether multiple layers developed during the simulations. The full impact of a core cutout on the global solution is unknown. A central cutout forces the use of additional boundary conditions at the shell's inner surface, potentially affecting flow morphology and transport properties.

In this study, we compute rotationally constrained solutions for semiconvection in the full sphere, including $r=0$ without the use of a central cutout. We focus on understanding the formation, morphology, and dynamics of semiconvective layers for different degrees of stratification and rotation. In Section~\ref{sec:numerical_model}, we describe the set of equations that model the hydrodynamics of semiconvection and the numerical methods. In Sections~\ref{sec:nr_results} and~\ref{sec:r_results}, we present the main results, first discussing non-rotating simulations, and then addressing the effects of rotation. Finally, we conclude in Section~\ref{sec:discussion} with a summary and discussion.

\section{Numerical Simulations}\label{sec:numerical_model}

In this section, we present the governing fluid equations and boundary conditions. We also describe the parameter space covered by the simulations and the initial conditions. We also briefly outline the numerical methods used.

\subsection{Model and Initial Conditions}
We consider a sphere of radius $R$ filled with a stably-stratified fluid comprised of two components, one light and the other relatively heavy (e.g., hydrogen and heavy elements). The concentration of the heavy component is represented by $C$. We adopt the Boussinesq approximation \citep{Spiegel_Veronis_1960}, i.e., we express the fluid density $\rho$ with a linear relationship between the temperature and composition, $\rho = \rho_0(1 - \alpha T + \beta C)$, where $\rho_0$ is a fiducial density and $\beta$ and $\alpha$ are the coefficients of compositional and thermal contraction/expansion (both assumed positive constants), respectively. Since the Boussinesq approximation is valid for subsonic flows and spatial scales smaller than a density scale height, our simulations should not be interpreted as models of entire stars or planets. Instead, they are more appropriate for the inner core of a gas giant, where the density varies by a factor of 2–4 within 20\%--25\% of the radius \citep[see, e.g.,][]{Helled_howard2024}. In gas giant interiors, the density scale height is comparable to the planetary radius, making the Boussinesq approximation reasonable for a small core. For stars, this assumption is more questionable, and we discuss its limitations later in the paper.  

We initialize the temperature and composition profiles to increase quadratically with depth, i.e., $T_0(r) = \Delta T_0 (1-r^2/R^2)$ and $C_0(r) = \Delta C_0 (1-r^2/R^2)$. By this choice the initial density ratio is constant across the sphere ($R_{\rho} = \beta \Delta C_0/\alpha \Delta T_0$).

We present the fluid equations in nondimensional form, using the radius of the sphere $R$ as the unit of length, and the thermal diffusion time $R^2/\kappa_T$ as the unit of time. For the units of temperature and composition, We use $[T] = \Delta T_0$ and  $[C] = \Delta C_0$, respectively. A unit of pressure corresponds to $[P] = \rho_0 (\kappa_T/R)^2$. Under this normalization, the surface of the sphere is located at $r=1$. The dimensionless equations are 

\begin{gather}
\nabla \cdot \bm{u} = 0\, , \label{eq:div u}\\
\nonumber \dfrac{\partial \bm{u}}{\partial t} + \bm{u}\cdot \nabla \bm{u} + \dfrac{\rm Pr}{\rm Ek} \bm{\hat{z}}\times \bm{u}  = - \nabla P + \mathrm{Ra}\mathrm{Pr}\left(R_\rho C - T\right)g(r)\bm{\hat{r}}\\
+ \mathrm{Pr} \nabla^2\bm{u} \, ,\\
\dfrac{\partial C}{\partial t} + \bm{u}\cdot \nabla C = \tau \nabla^2 C\, ,\\
\dfrac{\partial T}{\partial t} + \bm{u}\cdot \nabla T = \nabla^2 T\, , \label{eq:T}
\end{gather}
where $\bm{u}$ is the velocity field, and $g(r)= r$ is the dimensionless gravity. To be consistent with the initial profiles, and to ensure a stationary solution, we fix the gradient of temperature and composition at the surface of the sphere to match the initial conditions, so that $\partial_r T|_{r=1} = \partial_r C|_{r=1} = -2$~. The boundary conditions for the velocity are impenetrable and stress free, $u_r = \partial_r(u_\theta/r) =  \partial_r(u_\phi/r)=0$, where $u_r$, $u_{\theta}$ and $u_{\phi}$, the radial, polar, and azimuthal components of the velocity, respectively.

The set of equations \eqref{eq:div u}--\eqref{eq:T} is governed by five dimensionless numbers: the Prandtl number, the diffusivity ratio, and density ratio (defined in Equation~\ref{eq:parameters}), as well as the Rayleigh number and Ekman number, which are defined respectively as
\begin{gather}
    \mathrm{Ra} = \frac{\alpha g_0 \Delta T_0 R^3}{\kappa_T \nu} ,\hspace{0.3cm}
    \mathrm{Ek} = \frac{\nu}{2\Omega R^2}~,
\end{gather}
where $g_0$ is the gravity at $r=1$, and $\Omega$ is the rotation rate. 

All the simulations were run with the same Rayleigh number, Prandtl number, and diffusivity ratio (${\rm Ra} = 10^9$, ${\rm Pr} = 0.1$, and $\tau = 0.1$). Under this choice, double-diffusive instabilities are expected to arise for $ 1 < R_{\rho} < 5.5 $. Since we are interested in the layered regime, we varied the stability of the fluid by changing the density ratio $R_\rho$ from 1.25 up to 2. We set the Coriolis force to be identically zero in the non-rotating cases ($\mathrm{Ek} = \infty$), while for the rotating cases we vary the Ekman number between $\mathrm{Ek} \approx 2.6\times 10^{-6}$--$10^{-5}$. Real astrophysical conditions, where $\mathrm{Ra} \sim 10^{30}, \mathrm{Ek}\sim 10^{-20},~ \mathrm{Pr} \sim \tau \sim 10^{-7}$--$10^{-2}$ \citep{Jermyn2022} are inaccessible for current computational capabilities. However, by fixing $\mathrm{Ra}\gg 1$, $\mathrm{Ek} \ll 1$, $\mathrm{Pr} < 1$, and $\tau < 1$, we ensure that the simulations are qualitatively in the same dynamical regime as stars and giant planets.

\subsection{Simulation Details}

We time-evolve equations \eqref{eq:div u}--\eqref{eq:T} using the Dedalus pseudospectral solver \citep{Burns2020} version 3. The variables are represented in spherical harmonics for the angular directions and Jacobi polynomials for the radial direction. The number of radial, latitudinal, and longitudinal coefficients in all the simulations are $(N_r,N_\theta,N_\phi) = (256,384,768)$, respectively. For time-stepping, we use a second order semi-implicit BDF scheme \citep[SBDF2,][]{wang_ruuth_2008}, where the linear and nonlinear terms are treated implicitly and explicitly, respectively. To ensure numerical stability, the size of the time steps is set by the Courant–Friedrichs–Lewy (CFL) condition, using a safety factor of 0.2 (based on trial and error). To prevent aliasing errors, I apply the ``3/2 rule'' in all directions when evaluating nonlinear terms.
To start the simulations, we add random-noise temperature perturbations sampled from a normal distribution with a magnitude of $10^{-5}$ compared to the initial temperature field.

\section{Results for Non-rotating Simulations}\label{sec:nr_results}

In this section, we first present qualitative results of the evolution of the simulations for different density ratios. Then, we compare the transport properties with analytical predictions from previous studies. 

\subsection{Qualitative Evolution of the Fluid}

\begin{figure}
    \centering
    \includegraphics[width=\columnwidth]{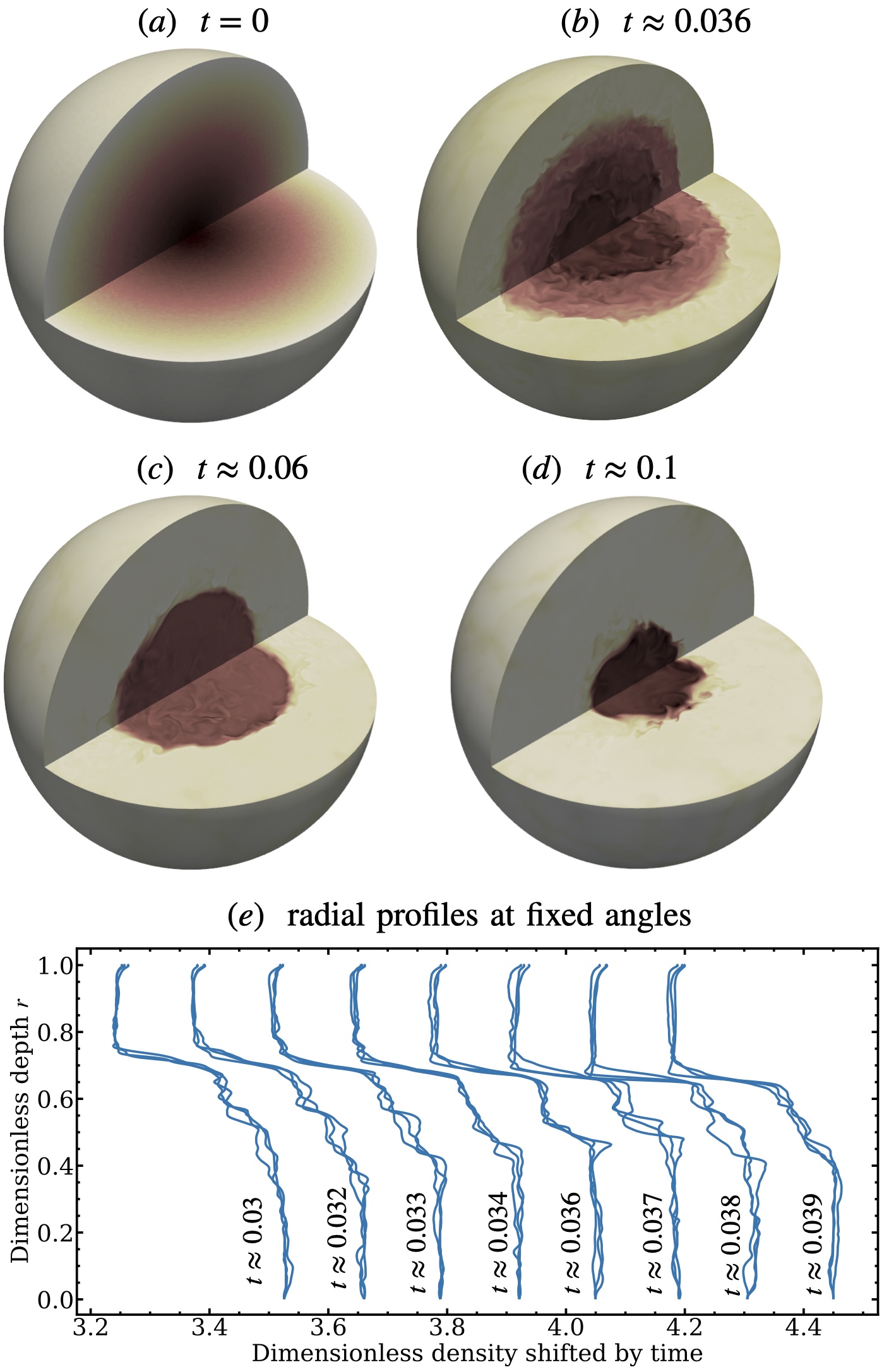}
\caption{Numerical results for the case $R_\rho = 1.25$. Panels (a)--(d): Snapshots of the density field at different times during the evolution of the system. Darker and lighter colors represent denser and less dense fluid, respectively. Panel (e): Radial density profiles for $t\approx 0.03$--0.04. As explained in the text, the superposition of different curves is due to radial profiles measured at different angles ($\phi=0$, $\theta = 0,~\pi/8,~\pi/4,~\pi/2,\pi$). The different stacks of density profiles are horizontally offset (shifted) from earlier time steps for ease of visualization.}
    \label{fig:non_rot_evolution}
\end{figure}

We begin with a reference case with stability parameter $R_{\rho} = 1.25$. This case is the closest to overturning convection.  We find that the fluid becomes rapidly unstable to semiconvection everywhere. There is a short phase where the system is in the regime of homogeneous oscillatory convection, with weak deviations from the initial density profile. Between $t \approx 0.03$--0.04, a few convective layers form spontaneously, with separating interfaces that are highly distorted due to overshooting motions. Since gravity (and in turn, the buoyancy fluxes) increases towards the surface, the system is dominated by a vigorous outer convection zone that propagates inwards, engulfing and mixing everything on its way (Figure~\ref{fig:non_rot_evolution}a--d). To clearly see the step-like structures of the layered regime of semiconvection (convective staircases), we show radial profiles of the density in Figure~\ref{fig:non_rot_evolution}e). Instead of performing a full spherical average, we measure the profiles at particular angles $\theta$ and $\phi$ because fluctuations due to overshooting motions cause an artificial interface broadening when horizontally-averaging over all $\theta$ and $\phi$. Although the profiles are noisy, they are consistent with the snapshot in panel (b), where 3 convective layers are clearly seen at $t\approx 0.036$. Finally, at $t\approx 0.12$, the outer convective layer reaches the center of the sphere and the entire fluid becomes fully-mixed.

\begin{figure}
    \centering
    \includegraphics[width=\columnwidth]{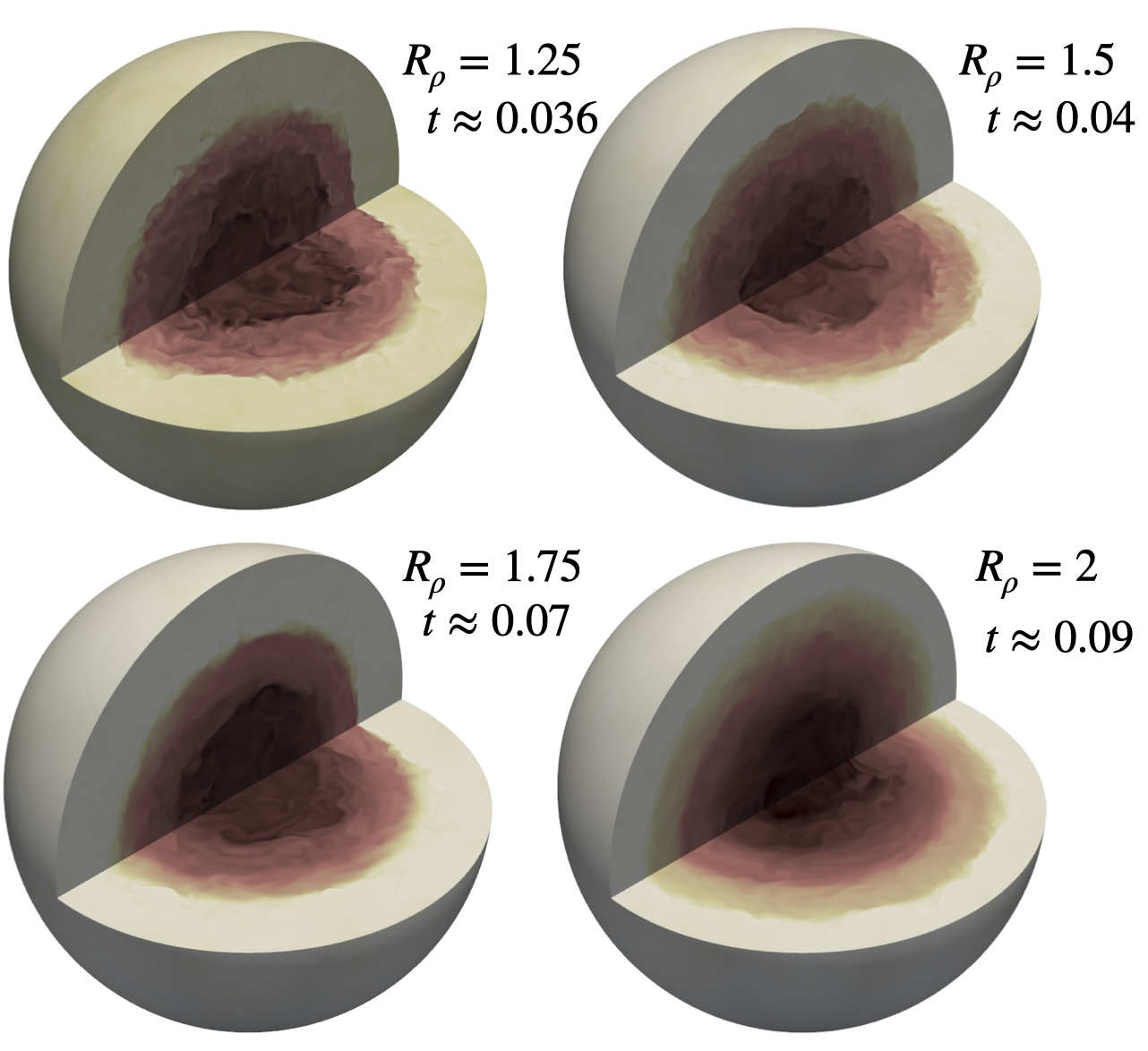}
\caption{Snapshots of the density field for simulations at different density ratios $R_\rho$. We selected snapshots at times where multiple convective layers were present in the fluid. Darker and lighter colors represent denser and less dense fluid, respectively.}
    \label{fig:non_rot_R_rho_snapshots}
\end{figure}

Simulations with larger density ratios ($R_\rho = 1.5,~1.75,2$) exhibit qualitatively similar behavior: multiple convective layers form in the fluid (Figure~\ref{fig:non_rot_R_rho_snapshots}), which subsequently merge and mix until the sphere becomes fully convective. However, the stability of the diffusive interfaces between the layers increases with larger $R_\rho$, as the stabilizing effect of the compositional gradient becomes stronger. Consequently, the timescale for the sphere to become fully mixed increases with $R_\rho$, with $t_{\rm{mix}} \approx 0.12,~0.175,~0.23,~0.32$ for $R_\rho = 1.25,~1.5,~1.75,~2$, respectively.

\subsection{Heat and Composition Transport}

We are interested in the radial fluxes of heat and composition through semiconvection. These fluxes are usually measured in terms of thermal and compositional Nusselt numbers, which are defined as the ratio of the total flux to the diffusive flux 
\begin{gather}
    \mathrm{Nu}_T  = 1 - \dfrac{\langle u_r T \rangle}{\langle dT/dr \rangle},~\\
    \mathrm{Nu}_C  = 1 - \dfrac{\langle u_r C \rangle}{\tau \langle dC/dr \rangle}~,
\end{gather}
where the notation $\langle \cdot \rangle$ represents a volume average over the entire sphere.

\begin{figure}
    \centering
    \includegraphics[width=\columnwidth]{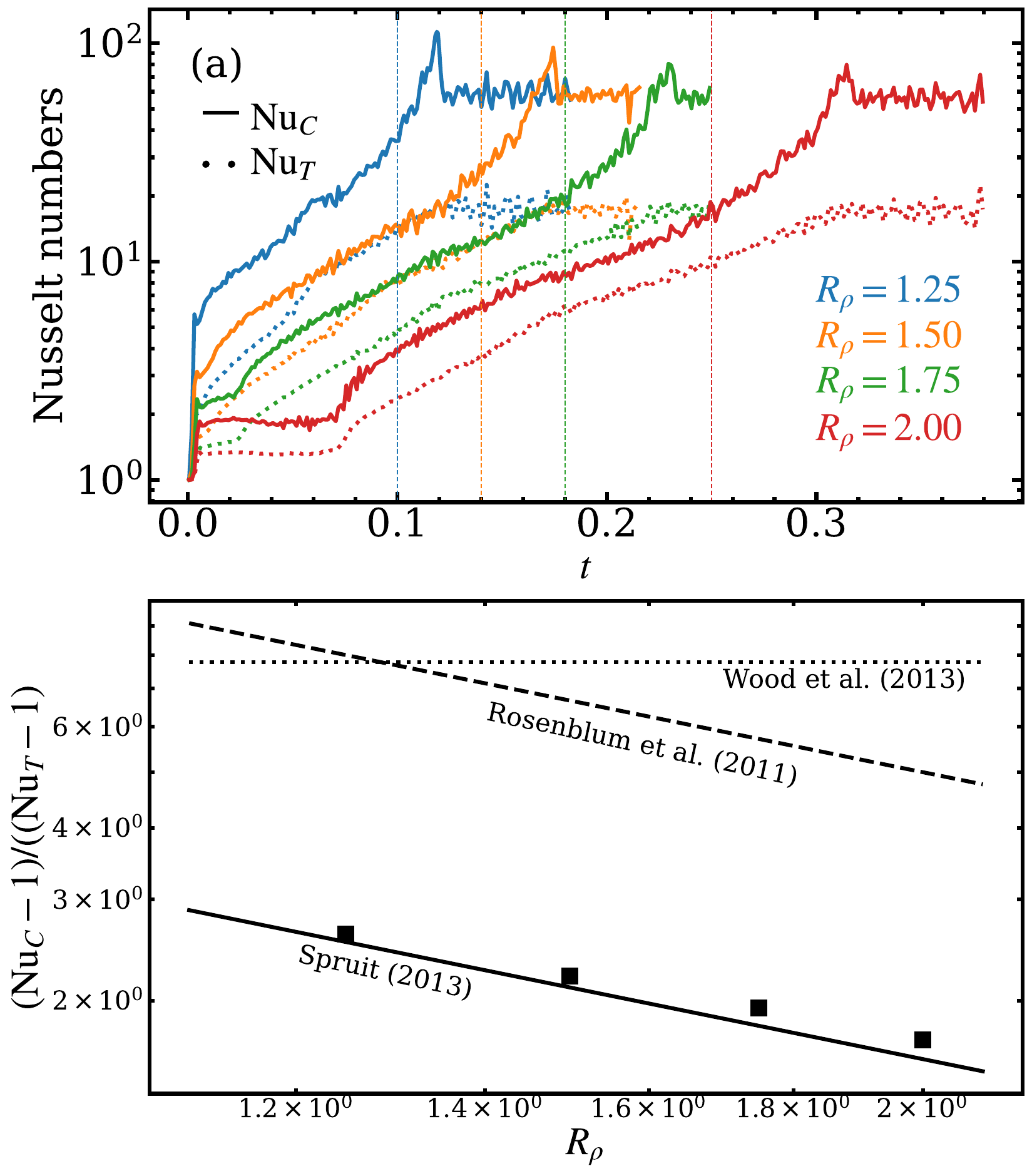}
    \caption{Panel (a): volume averages of the compositional Nusselt number  $\mathrm{Nu_C}$ and  thermal Nusselt number $\mathrm{Nu_T}$, as a function of time. The vertical lines correspond to the times at which the dynamics of the flow is dominated by the outer convection zone. Results are shown for simulations at different density ratios $R_{\rho}$. Panel (b): $(\mathrm{Nu}_C-1)/(\mathrm{Nu}_T-1)$ averaged over the semiconvective phase (see text), as a function of density ratio $R_\rho$. The lines correspond to theoretical predictions by \cite{Spruit2013} (Equation~\ref{eq:spruit}), \cite{Rosenblum2011} (Equation~\ref{eq:rosenblum}), and \cite{Wood2013} (Equation~\ref{eq:wood}).}
    \label{fig:nu_no_rot}
\end{figure}

Figure~\ref{fig:nu_no_rot}(a) shows time series of the thermal and compositional Nusselt numbers during the evolution of the simulations. In all cases, the Nusselt numbers increase during different phases characterized by different slopes of the curves with time. First, there is an exponential increase where the semiconvective instability grows and saturates once nonlinear effects become important. Once semiconvection fully establishes and layers form and evolve, both numbers increase over time but a smaller rate. Later on (starting from the times denoted by the vertical lines in Figure~\ref{fig:nu_no_rot}a), the increase in the turbulent transport becomes steeper with time as the fluxes become dominated by the more vigorous outer convection zone. Note that due to the reduced overall stability of the flow at lower $R_\rho$, the contribution from convective fluxes and consequently the magnitude of $\mathrm{Nu}$, is higher for smaller $R_\rho$.  Finally, as the entire sphere transitions to a fully convective and well-mixed state, the Nusselt numbers reach a peak and settle to a constant value, independent of $R_\rho$, but controlled by the imposed fluxes at the surface of the sphere which are the same for all the simulations. 

There exist different models for the relationship between the thermal and compositional Nusselt numbers. For example, in the analytical model of \cite{Spruit2013}, the Nusselt numbers are related via
\begin{gather}
\dfrac{\mathrm{Nu}_C - 1}{\mathrm{Nu}_T - 1} = \dfrac{q}{R_\rho \tau^{1/2}},~ \label{eq:spruit}
\end{gather}
where $q$ is a constant of order unity.

On the other hand, the numerical simulations of \cite{Rosenblum2011} and \cite{Wood2013} propose
\begin{gather}
\dfrac{\mathrm{Nu}_C - 1}{\mathrm{Nu}_T - 1} = \dfrac{1}{R_\rho \tau},~\label{eq:rosenblum}\\
\dfrac{\mathrm{Nu}_C - 1}{\mathrm{Nu}_T - 1} = \dfrac{B}{A}\dfrac{\mathrm{Ra}^{0.37-1/3}}{\mathrm{Pr}^{1/12} \tau},~\label{eq:wood}
\end{gather}
respectively, where the prefactors $A\approx 0.1$ and $B\approx 0.03$ are weakly dependent on $R_{\rho}$, $\mathrm{Pr}$, $\tau$, or $\mathrm{Ra}$.  

We tested the three models against measurements of the Nusselt numbers from the simulation results. When averaging over the semiconvective phase (i.e. the time span between the end of the initial exponential grow and the vertical lines in Figure~\ref{fig:nu_no_rot}(a)), we find that none of the models give the exact measured values from the simulations. However, Spruit's theoretical prediction (Equation~\ref{eq:spruit}) gives a better fit to the data (Figure~\ref{fig:nu_no_rot}(b)). Note that Rosenblum's prediction  (Equation~\ref{eq:rosenblum}) cannot be ruled out, as it provides the correct scaling with the density ratio $R_\rho$, and adjusting the prefactor to $\approx 0.33$ also gives a good fit.

\begin{figure}[h!]
    \centering
    \includegraphics[width=\columnwidth]{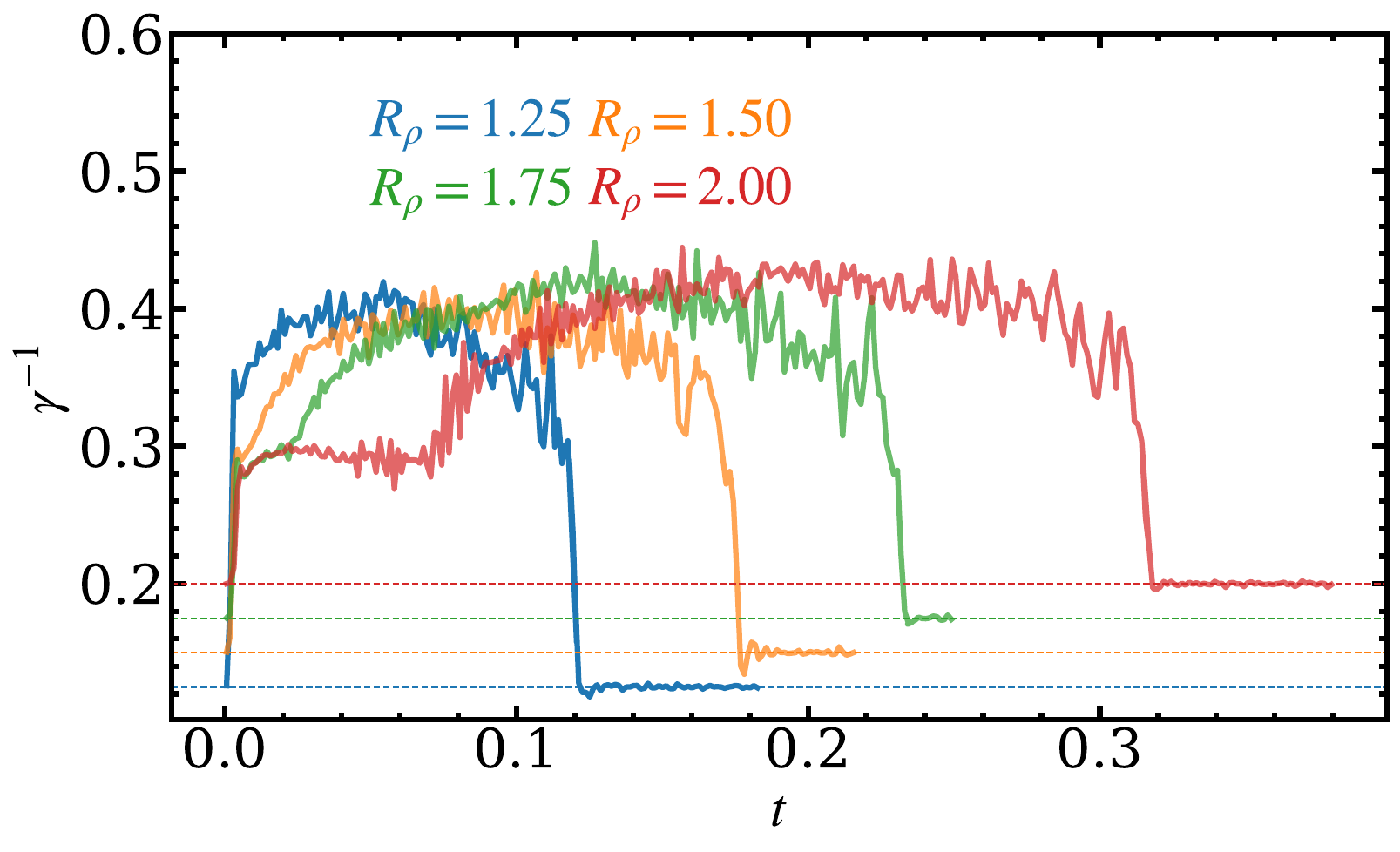}
    \caption{Time series of the buoyancy flux ratio $\gamma^{-1}$ (Equation~\ref{eq:gamma}) for different density ratios $R_\rho$. The dashed line correspond to $\tau R_{\rho}$, i.e., the expected volume averaged $\gamma^{-1}$ when the entire sphere becomes fully mixed and its temperature and composition decrease everywhere at a constant rate.}
    \label{fig:gamma_no_rot}
\end{figure}

Another important quantity in recent studies of semiconvection is the ``buoyancy flux ratio'' $\gamma^{-1}$, defined as the ratio of the buoyancy flux associated with compositional
transport to the equivalent buoyancy flux
associated with heat transport \citep[see, e.g.,][]{Garaud2018,Garaud2021}

\begin{gather}
\gamma^{-1} = R_\rho \dfrac{\langle u_r C - \tau dC/dr\rangle }{\langle u_r T - dT/dr\rangle }~. \label{eq:gamma}
\end{gather}
As the sphere becomes mixed over time, the gravitational potential energy released from the thermal stratification must exceed the potential energy required to mix the compositional stratification, implying $\gamma^{-1} < 1$. Figure~\ref{fig:gamma_no_rot} shows that $\gamma^{-1}$ remains below unity these simulations. Initially, during the homogeneous phase (before layer formation), $\gamma^{-1}$ decreases as a function of $R_\rho$. Over time, $\gamma^{-1}$ increases and reaches a peak, although no clear trend emerges for how its maximum value varies with $R_\rho$. Previous studies have proposed that $\gamma^{-1} = \tau^{1/2}$ in experiments of semiconvection in salty water \citep[e.g.,][]{Turner_1965,Linden1975}, or $\gamma^{-1} \sim \tau (\mathrm{Pr} + 1)/(\mathrm{Pr} + \tau)$ in numerical simulations of semiconvection in triply periodic boxes \citep[e.g.,][]{Moll2017}. For $\tau = \mathrm{Pr} = 0.1$ these predictions yield $\gamma^{-1} \approx 0.32$ and $0.55$, respectively. In comparison, our results range from 0.3 to 0.4 during the semiconvective regime. Once the sphere becomes fully mixed, both the composition and heat flux reach a statistically steady state, varying linearly with radius. At this stage, the volume-averaged value of $\gamma^{-1}$ approaches the steady-state solution, $\tau R_\rho$ (horizontal lines in Figure~\ref{fig:gamma_no_rot}). On average, we find that our results are broadly consistent with those from non-rotating simulations in Cartesian boxes \citep[see, e.g., ][for a review]{Garaud2021}.

\section{Results for Rotating Simulations}\label{sec:r_results}


We analyze the rotating runs using the Rossby number, $\mathrm{Ro}$, of the flow. This nondimensional
number expresses the ratio of rotational to convective
timescales,  so that a system subject to significant Coriolis force possesses low $\mathrm{Ro}$. Convection that is relatively insensitive to rotation is characterized by a high value of $\mathrm{Ro}$. While the Rossby number can be computed
using the characteristic speed and length scale of the resultant flow, it can also be estimated a priori by system control parameters, effectively using the freefall time across the domain $t_{\mathrm{ff}} =  R/U_{\mathrm{ff}}$, where $U_{\mathrm{ff}} = (\alpha g\Delta T_0 R)^{1/2}$, as a proxy for the convective timescale

\begin{gather}
\mathrm{Ro_c} = \dfrac{U_{\mathrm{ff}}}{2\Omega R} = \left(\dfrac{\alpha g \Delta T_0 R}{4\Omega^2 R^2}\right)^{1/2} = \left(\dfrac{\mathrm{Ra}}{\mathrm{Pr}}\right)^{1/2}\mathrm{Ek}~. \label{eq:conv_Ro}
\end{gather}
This definition is known as the convective Rossby number. Our rotating simulations have $R_{\rho} = 1.25$ and $\mathrm{Ro_c} = 0.25,~ 0.5,  ~1$. 

\subsection{Qualitative Evolution and Flow Morphology}

\begin{figure}
    \centering
    \includegraphics[width=\columnwidth]{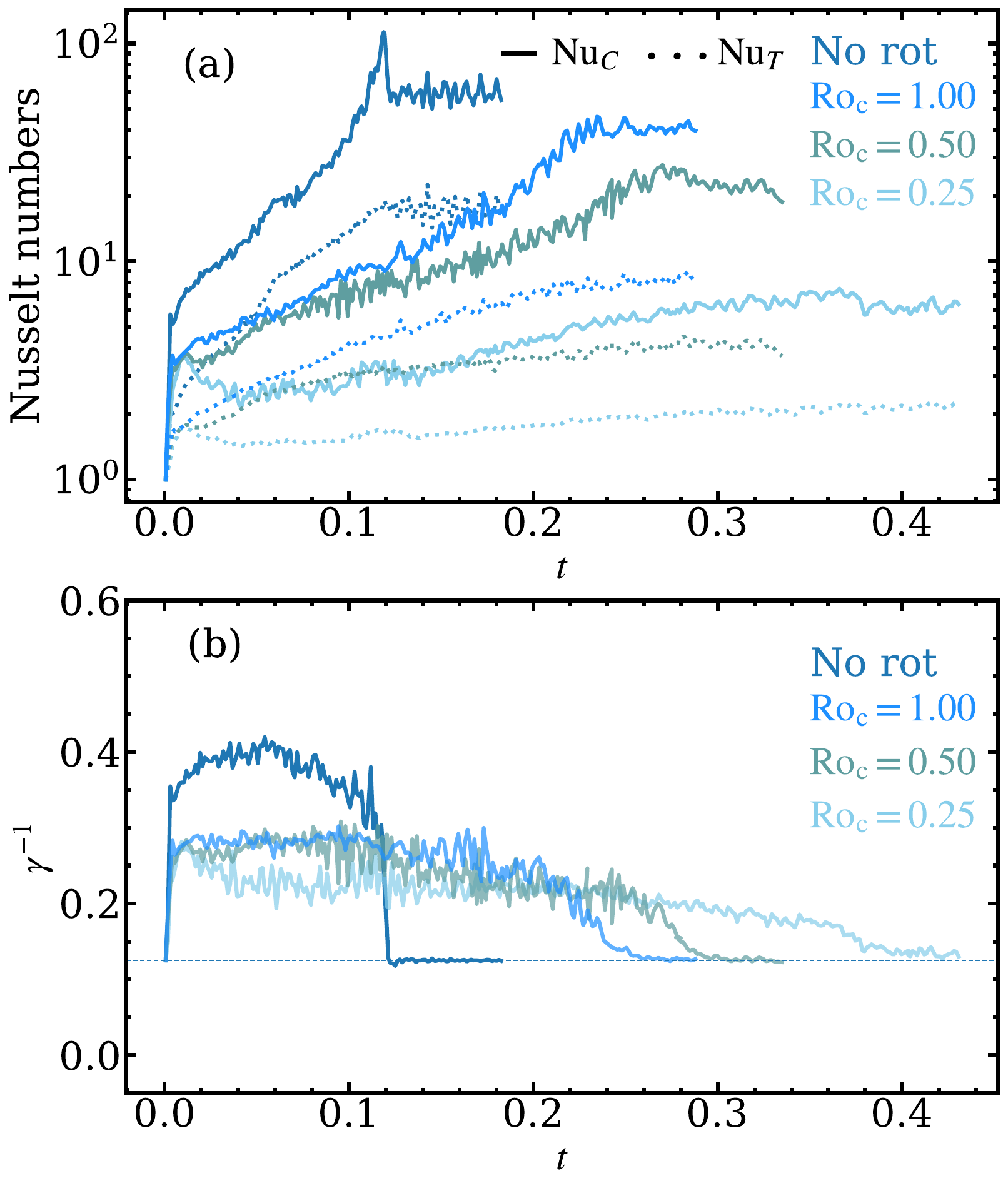}
    \caption{Panel (a): volume averages of the compositional Nusselt number  $\mathrm{Nu_C}$ and  thermal Nusselt number $\mathrm{Nu_T}$, as a function of time. Results are shown for simulations at different convective Rossby numbers (0.25, 0.5, and 1) but fixed $R_{\rho} = 1.25$. For comparison, we include the nonrotating case. Panel (b): time series of the buoyancy flux ratio $\gamma^{-1}$ (Equation~\ref{eq:gamma}) for the same cases in panel (a). The dashed line correspond to $\tau R_{\rho}$, i.e., the expected volume averaged $\gamma^{-1}$ when the entire sphere becomes fully mixed and its temperature and composition changes at a constant rate.}
    \label{fig:nu_gamma_rot}
\end{figure}

The evolution of the rotating runs is qualitatively similar to that of the nonrotating cases, i.e., once the semiconvective instability is triggered, a few convective layers form, which grow and merge until the entire sphere is well-mixed and fully convective. However, there are some significant differences with respect to the nonrotating counterpart. First, the thermal and compositional transport measured through the Nusselt numbers and the buoyancy flux ratio are much smaller than for the nonrotating case (see Figure~\ref{fig:nu_gamma_rot}). Consequently, the sphere becomes fully mixed at later times, with the difference respect to the nonrotating case being greater as the Rossby number decreases ($t_{\mathrm{mix}} \approx 0.4,~ 0.3,~0.26$ for $\mathrm{Ro_c} = 0.25,~0.5,~1$, respectively). Further, unlike the nonrotating cases, once the sphere becomes fully convective, the Nusselt numbers are not the same  between the different simulations, despite that $R_\rho$ and the surface gradients are equal in all the simulations. The explanation for this discrepancy is the weaker convective transport and the well-known enhancement of the thermal gradient in rotating convection, which becomes more pronounced as the Rossby number decreases  \citep[see, e.g.,][]{Barker2014}. This implies that, compared to the non-rotating case, a larger fraction of the heat flux is carried by conduction rather than convection as rotation increases. Overall, rotation exhibits a stabilizing effect on convective fluxes, similar to that of increasing the stability ratio $R_\rho$ in nonrotating simulations. Note since we run simulations at fixed $R_\rho = 1.25$, we cannot properly test the analytical predictions in Equations~\ref{eq:rosenblum}--\ref{eq:wood}. However, we found that the ratio between the Nusselt numbers varies between 2.5 and 3.5, i.e., close to that of the nonrotating simulation at the same density ratio.

\begin{figure*}
    \centering
    \includegraphics[width=0.9\textwidth]{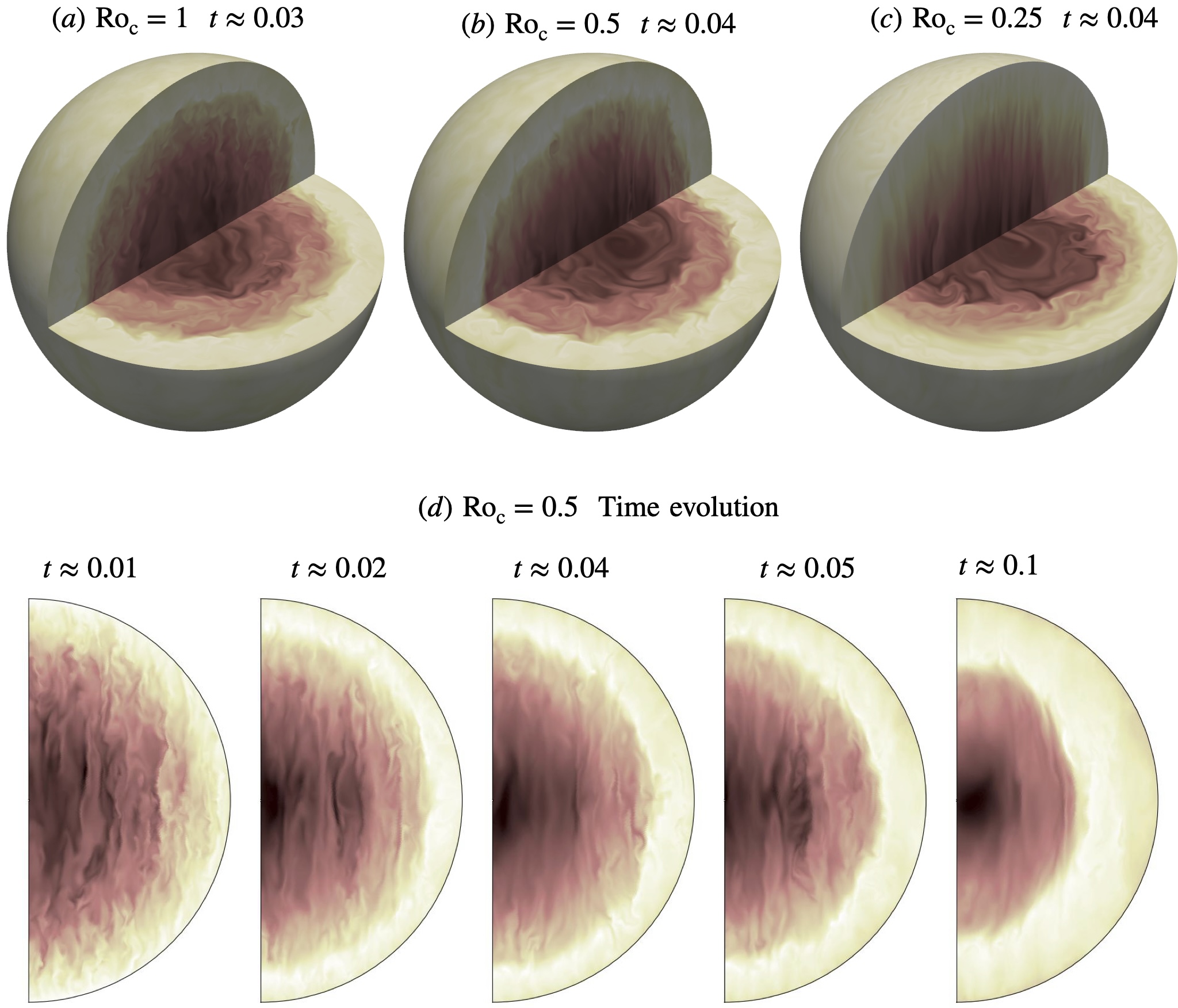}
    \caption{Meridional snapshots of the density field for simulations at fixed $R_\rho = 1.25$ but different convective Rossby numbers (1, 0.5 and 0.25 in panels a, b, and c, respectively). We selected snapshots at times where a few convective layers were present in the fluid. Panel (d): Meridional snapshots of the density field at different times for the case at $\mathrm{Ro_c} = 0.5$. In all panels, darker and lighter colors represent denser and less dense fluid, respectively}
    \label{fig:flow_rot}
\end{figure*}

Another difference is that rotation significantly affects the spatial scales of the convective flow. In the nonrotating simulations, the flow exhibits approximately isotropic length scales, whereas in the rotating cases, the horizontal length scale (perpendicular to the rotation vector) becomes much smaller than the vertical scale as rotation increases. This behavior is consistent with the Taylor-Proudman theorem, which predicts the formation of vortices and thin columns aligned with the rotation axis \citep[e.g.,][]{Proudman1916,Taylor1917}. This difference is clearly seen for the case at $\mathrm{Ro_c} = 0.25$,  at mid and high latitudes (Figure~\ref{fig:flow_rot}c). 

To illustrate how the fluid mixes over time, we show in Figure~\ref{fig:flow_rot}(d) meridional planes of the density field for $\mathrm{Ro_c} = 0.5$. We selected this case because it has an intermediate level of rotational influence, and evolves in a similar way to $\mathrm{Ro_c} = 0.25$, whereas the case $\mathrm{Ro_c} = 1$ behaves more similar to the nonrotating cases. Initially, the flow is dominated by elongated structures aligned with the rotation axis, which subsequently merge to form 3--4 convective layers. The lack of isotropic density redistribution, which would be characteristic of spherical mixing, further emphasizes the role of rotational effects in organizing the flow. For example, the extent of mixing in the equatorial regions is clearly larger than at the poles. This indicates that mixing is governed by cylindrical geometry, shaped by the interaction between buoyancy and Coriolis forces.

We also observe the formation of prograde zonal flows at the surface, which alternate and reverse to retrograde and then back to prograde as a function of radius. The radial shear between the flows significantly distorts the interfaces between the layers. A comprehensive study exploring the fluid's transport properties, flow morphology, and zonal flow characteristics as a function of $\mathrm{Ra}$, $R_\rho$, $\mathrm{Pr}$, and $\tau$ will be presented in a future publication. 

\subsection{Latitudinal Dependence of the Heat and Composition Transport}

Given the significant differences between convective flows in the polar regions and those at lower latitudes near the equator, we quantify these variations by measuring local Nusselt numbers as a function of latitude. Since we aim to compare with results from previous studies of thermal convection in rotating shells, we show azimuthally averaged profiles of the Nusselt numbers at $r\approx 0.9$, i.e., in the outer convective layer where the fluxes and the velocities are larger. To ensure the flux measurements are captured early in the semiconvective phase and to further reduce noise, we time-average the profiles over $\Delta t \approx 0.01$ centered around $t\approx 0.05$. 

\begin{figure}
    \centering
    \includegraphics[width=\columnwidth]{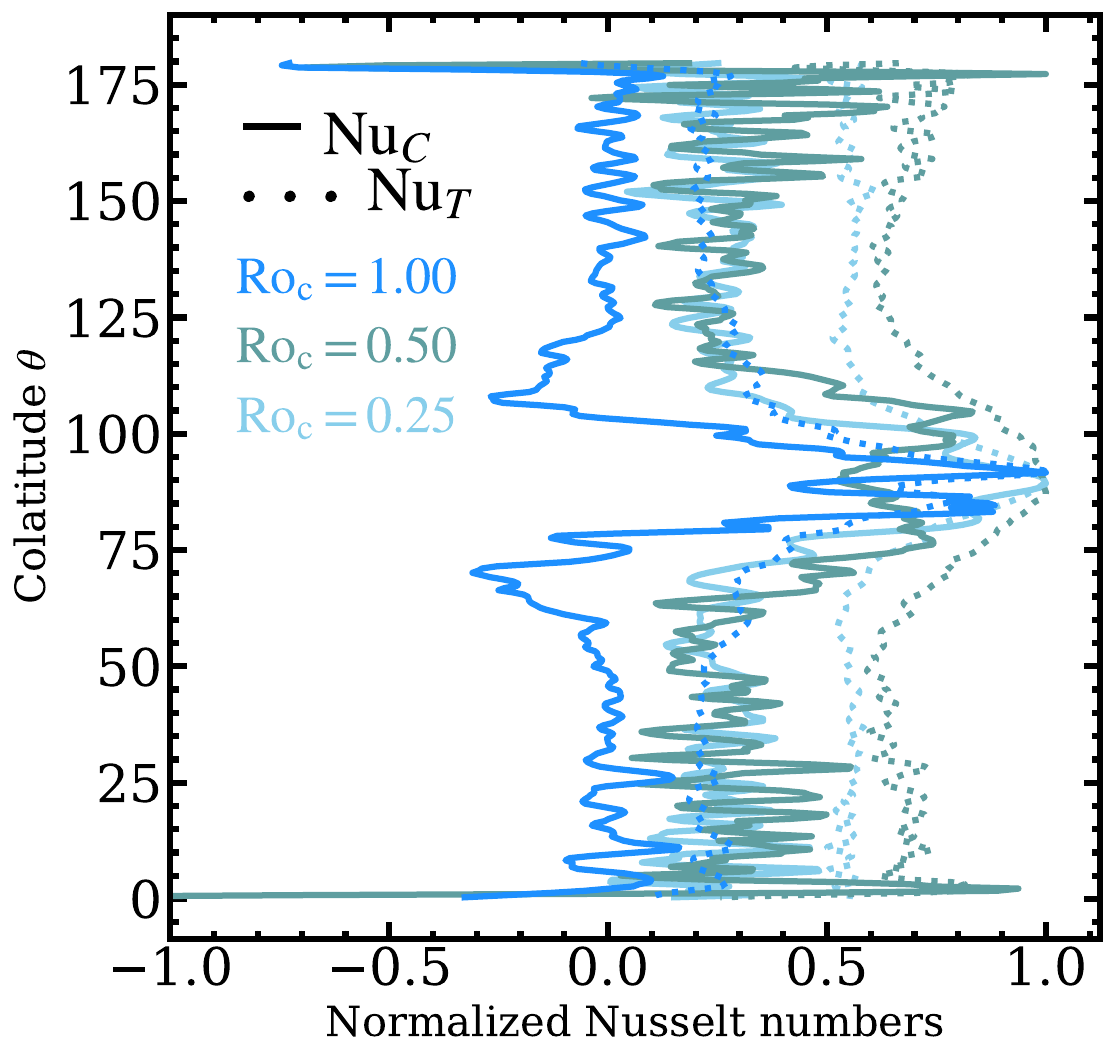}
    \caption{Time-averaged local Nusselt numbers normalized by their maximum values (so that they are 1 at the equator) as a function of the colatitude for simulations at $\mathrm{Ro_c} = 0.25$, 0.5, and 1. The results are shown within the outer convection zone, at $r=0.9$, where the fluxes are stronger. The solid and dotted lines distinguish between the compositional and thermal Nusselt numbers, respectively. Similarly to thermal convection in rotating spherical shells, the radial transport of both heat and composition is more efficient at the equator.}
    \label{fig:Nu_profiles}
\end{figure}

Figure~\ref{fig:Nu_profiles} shows that both compositional and heat transport in the radial direction are more effective near the equator while becoming less efficient at higher latitudes. This aligns well with results from previous studies focusing on thermal convection alone \citep[see, e.g., Figure~3c in][]{Wang2021}. Recently, \cite{Gastine2023} showed that differences in heat transport as a function of latitude depend strongly on the supercriticality of the flow, defined as $\mathcal{R} = \mathrm{Ra}\mathrm{Ek}^{4/3}$. For $\mathcal{R} \sim 1$--$10$, convective transport is expected to be more efficient at the equator than at the poles, leading to larger Nusselt numbers at the equator. However, for larger supercriticalities ($\mathcal{R} > 10$), heat transport at the poles becomes increasingly efficient as $\mathcal{R}$ increases, reaching a point where the equator and poles transport heat equally efficiently for $\mathcal{R} > 100$, when rotational effects become less influential. The Nusselt numbers in Figure~\ref{fig:Nu_profiles} are measured at times when the fluid is not fully convective, leading to $\mathcal{R} \sim 10$--$100$. This means that our results align well with the regime of more efficient convection at the equator.

\section{Discussion}\label{sec:discussion}

\subsection{Summary of the Main Results}

In this work, we have studied the onset and evolution of semiconvection in astrophysical fluids, i.e., in fluids of small Prandtl number $\mathrm{Pr}=\nu/\kappa_T$, and small diffusivity ratio $\tau = \kappa_C/\kappa_T$. Unlike previous work which was limited to conducting simulations on Cartesian boxes, or spherical shells imposing artificial boundary conditions on the inner surface and potentially altering the dynamics, we have solved the Boussinesq fluid equations for the entire sphere, including $r = 0$, for the first time, and including the Coriolis force. For our choice of $\mathrm{Pr} = \tau = 0.1$, we varied the density ratio $R_\rho$ between 1.25 and 2 for the non-rotating runs, and we fixed $R_\rho = 1.25$ while varying the convective Rossby number $\mathrm{Ro_c}$ between 0.25 and 1 in the rotating cases.

In all cases, we observed an initial phase of weak, instability-driven turbulence, followed by the formation of a few convective layers in the fluid (see Figures~\ref{fig:non_rot_evolution}, \ref{fig:non_rot_R_rho_snapshots}, and \ref{fig:flow_rot}). These layers quickly begin to merge until the sphere becomes fully mixed, with the mixing time increasing with $R_\rho$ due to the enhanced stability of the composition gradient, and also increasing for smaller convective Rossby numbers $\mathrm{Ro_c}$ due to the much smaller heat and compositional transport hampered by rotation (see measurements of the radial Nusselt numbers in Figures~\ref{fig:nu_no_rot}(a) and \ref{fig:nu_gamma_rot}(a)). We observe a similar trend during the semiconvective phase (i.e. before the sphere becomes fully convective and well mixed) when comparing the buoyancy flux ratio $\gamma^{-1}$ for different $R_\rho$ and $\mathrm{Ro_c}$ (see Figures~\ref{fig:gamma_no_rot} and \ref{fig:nu_gamma_rot}b). Interestingly, when testing our measurements of the radial Nusselt numbers during the semiconvective phase against theoretical predictions from previous work \citep{Rosenblum2011,Spruit2013,Wood2013}, we found that Spruit's theory (Equation~\ref{eq:spruit}) gives a better agreement with the data. However, Rosenblum's prediction (Equation~\ref{eq:rosenblum}) gives the right scaling with the density ratio ($[\mathrm{Nu}_C-1]/[\mathrm{Nu}_T-1] \propto 1/R_\rho$), matching the data with a different prefactor equal to 0.33. Therefore, more work is needed to better distinguish between the two models, especially if the prefactor depends on the boundary conditions or geometry of the system.

A noticeable difference between the nonrotating and rotating runs is the morphology of the convective flow. In the nonrotating case, convection exhibits nearly isotropic spatial scales, whereas in the rapidly rotating cases, the flows become anisotropic, with longer axial scales compared to horizontal scales (see Figures~\ref{fig:non_rot_R_rho_snapshots} and \ref{fig:flow_rot}). These differences are also reflected in the radial transport as a function of colatitude (Figure~\ref{fig:Nu_profiles}). Near the poles, the radial component of the convective flow is weaker due to the columnar morphology of the fluid, which favors horizontal fluid motion. In contrast, radial transport is more efficient at the equator because the Coriolis force does not significant affect radial motions.

\subsection{Comparison to Previous Work}

Semiconvection in our non-rotating simulations in the full sphere behaves qualitatively similar to previous work on non-rotating Cartesian boxes, with only minor differences with respect to the critical density ratio at which layers form. For example, in the suite of simulations by \cite{Mirouh2012}, for $\mathrm{Pr}=\tau=0.1$, layer formation occurs when $R_\rho \approx 1.5$. In these simulations, we find layer formation even for $R_\rho = 2$. The fluxes measured through the buoyancy flux ratio are similar. For example, \cite{Mirouh2012} measured $\gamma^{-1} \approx 0.3$--0.5 for the same values of $R_\rho$, Pr, and $\tau$ considered in this work (see their Table 5). We measured $\gamma^{-1} \approx$ 0.3--0.4 during the semiconvective phase. A more significant difference is in the Nusselt numbers, where \cite{Mirouh2012} measured $\mathrm{Nu}_C \sim 10$,  $\mathrm{Nu}_T \sim 1$--5, while we measured $\mathrm{Nu}_C \sim 100$,  $\mathrm{Nu}_T \sim 10$. These differences are not surprising given the significant differences in simulation setups. Our simulations were conducted on a full spherical domain, with gravity and fluxes increasing toward the surface. In contrast, the simulations in \cite{Mirouh2012} were performed in Cartesian boxes with constant gravity, triply periodic boundary conditions, and no external flux forcing beyond the initial profiles. However, we emphasize that the formation and evolution of the system are qualitatively similar, with no significant differences. 

Comparison with previous work that includes rotation is more difficult because very little is known about rotating semiconvection. \cite{Blies2014} investigated semiconvection in rotating spherical shells, but it remains unclear whether they observed layer formation. The Rayleigh number used in their simulations was $\sim 10^7$, which corresponds to a domain size of about $50d$, where $d\approx (\kappa\nu/\alpha g |dT/dz|)^{1/4}$ is the characteristic scale of double-diffusive eddies. Typically, in simulations of semiconvection, the first layers that form are never thinner than 30$d$--50$d$ \citep{Garaud2021}. So, it is likely that \cite{Blies2014} did not find layers because of the smaller Rayleigh number (our simulations have $\mathrm{Ra} = 10^9$, which corresponds to $\sim 180d$, so that a few layers are expected to form). Nevertheless, consistent with the findings of this study, they reported that rotation reduces both heat and compositional fluxes. When testing the theoretical predictions described earlier, they also found that Spruit's theory provides a better approximation to the data. 

An intriguing suite of simulations was conducted by \cite{Moll2017_rot} on Cartesian boxes. Similar to the results presented here, they demonstrated that rotation consistently diminishes thermal and compositional transport across the system. Interestingly, in some of their simulations, where the flow was strongly constrained by rotation (low Rossby numbers), layer formation was suppressed, and large-scale vortices formed instead, enhancing compositional transport. However, their results appear to be highly sensitive to the size of the boxes adopted in the simulations, likely due to their use of triply periodic boundary conditions. More recently, \cite{Fuentes2024} conducted 3D simulations on Cartesian boxes but using fixed flux boundary conditions. They demonstrated that rotation reduces the kinetic energy flux available for mixing, and, as consequence, rotation prolongs the lifetime of semiconvective layers.

Finally, the latitudinal dependence of the fluxes in our simulations aligns well with previous studies of thermal convection in rapidly rotating shells \citep[e.g.,][]{Wang2021,Gastine2023}.

\subsection{Limitations and Future Work}

We have made a number of approximations that must be relaxed before making conclusions for stars and planets. First, we limited the simulations to fluids of $\mathrm{Pr}=\tau = 0.1$. These values are at the middle of values expected to occur in planetary interiors, where the Prandlt number and diffusivity ratio may extend down to $\sim 0.01$ \citep[e.g.,][]{Stevenson1977,French2012}. In stellar interiors, even lower values are expected. For example, in non-degenerate regions of MS and RGB stars, $\mathrm{Pr} \sim 10^{-6}$ and $\tau \sim 10^{-7}$ \citep[e.g.,][]{Garaud2015}. These parameter values are beyond current computational capabilities.

Second, we adopted the Boussinesq approximation, which restricts the vertical scale to be much smaller than a density scale height. Incorporating density stratification would change the mixing rate of the layers through a combination of two effects. First, the potential energy needed for mixing would decrease. While composition is uniformly mixed within a convective layer in both cases, stratified convection adjusts the density to the adiabatic gradient, whereas Boussinesq convection maintains a constant density, which requires more energy. Second, stratification enhances the asymmetry between upflows and downflows \citep{Meakin2007}, thereby increasing the net kinetic energy flux available for mixing. At present, we cannot speculate on the net impact of these effects, since, to the best of our knowledge, there are no simulations of semiconvection incorporating density stratification.

Third, we imposed fixed temperature and composition gradients at the surface of the sphere, which, despite that do not affect the layer formation process, are certainly not realistic boundary conditions for stars or planets. For example, in giant planets, the surface flux decreases over time, and there are no composition fluxes at the surface. In stars, internal heat generation and compositional changes from nuclear burning play a significant role, both of which are not included in these simulations. 

Improvements in numerical modeling, focusing on adding density stratification, and the use of more suitable boundary conditions, would be of great interest to inform 1D evolution models and interpret the rich set of observations of stars and planets.

\begin{acknowledgements}
I am grateful to Andrew Cumming, Bradley Hindman, and Pascale Garaud for many interesting conversations and valuable suggestions. I am also grateful to the Institute for Pure and Applied Mathematics (IPAM) at UCLA, for support and
hospitality during the workshop ``Rotating Turbulence: Interplay and Separability of Bulk and Boundary Dynamics'' supported by NSF grant DMS-1925919. This research was supported by NASA Solar System Workings grant 80NSSC24K0927.
\end{acknowledgements}

\bibliography{references}{}
\bibliographystyle{aasjournal}

\end{document}